\documentclass[twocolumn,tighten]{aastex631}

\hypersetup{linkcolor=red,citecolor=blue,filecolor=cyan,urlcolor=magenta}

\newcommand{\Chandra}{{\em Chandra}}
\newcommand{\lum}{erg s$^{-1}$}
\newcommand{\flux}{erg s$^{-1}$ cm$^{-2}$}
\newcommand{\Msun}{$M_{\odot}$}

\newcommand{\Zsun}{$Z_{\odot}$}

\newcommand{\Lx}{$L_X$}

\shorttitle{HMXB Variability and XLF in IC 10}
\shortauthors{Binder et al.}

\begin{document}

\title{The X-ray Variability and Luminosity Function of High Mass X-ray Binaries in the Dwarf Starburst Galaxy IC 10}

\correspondingauthor{Breanna A. Binder}
\email{babinder@cpp.edu}

\author[0000-0002-4955-0471]{Breanna A. Binder}
\affiliation{Department of Physics and Astronomy, California State Polytechnic University Pomona, CA, 91768, USA}

\author[0000-0001-6383-2777]{Rosalie Lazarus}
\affiliation{Department of Physics and Astronomy, California State Polytechnic University Pomona, CA, 91768, USA}

\author{Mina Thoresen}
\affiliation{Department of Physics and Astronomy, California State Polytechnic University Pomona, CA, 91768, USA}

\author{Silas Laycock}
\affiliation{Department of Physics and Applied Physics, University of Massachusetts Lowell, MA, 01854, USA}

\author{Sayantan Bhattacharya}
\affiliation{Department of Astronomy and Astrophysics, Tata Institute of Fundamental Research, Mumbai, 400005, India}

\begin{abstract}

We present an analysis of $\sim$235 ks of \Chandra\ observations obtained over $\sim$19 years of the nearby dwarf starburst galaxy IC~10 in order to study the X-ray variability and X-ray luminosity function (XLF) of its X-ray binary (XRB) population. We identify 23 likely XRBs within the 2MASS $K_S$ isophotal radius and find the distributions of their dynamic ranges and duty cycles are consistent with a young, high-mass XRB population dominated by supergiant-fed systems, consistent with previous work. In general, we find that brighter HMXBs (those with \Lx$\gtrsim$several$\times10^{36}$ \lum) have higher duty cycles (i.e., are more persistent X-ray sources) than fainter objects, and the dynamic ranges of the sgHMXBs in the lower metallicity environment of IC~10 are higher than what is observed for comparable systems in the Milky Way. After filtering out foreground stars on the basis of {\em Gaia} parallaxes we construct, for the first time, the XLF of IC~10. We then use the XLF to model the star formation history of the galaxy, finding that a very recent  (3-8 Myr) burst of star formation with rate of $\sim$0.5 \Msun\ yr$^{-1}$ is needed to adequately explain the observed bright-end (\Lx$\sim10^{37}$ \lum) of the HMXB XLF.

\end{abstract}

\keywords{High mass X-ray binary stars (733) --- X-ray transient sources (1852) --- Starburst galaxy (1570) --- Dwarf galaxy (416)}

\section{Introduction} \label{sec:intro}
Nearby galaxies provide our most detailed views of the processes that dominate the evolution of a wide variety of galaxy types, masses, and metallicities, while containing rich reservoirs of every stage of stellar evolution. Accretion-powered X-ray binaries (XRBs) containing black holes (BHs) and neutron stars (NSs) are the relics of the most massive and short-lived binary stars; their current properties result from the orbital and mass-exchange history of the progenitor binary, and they ultimately form millisecond pulsars, short gamma-ray bursts, or double compact object systems. Feedback from XRBs, especially high-mass XRBs (HMXBs) is now recognized as an important regulator of baryonic mass in early galaxies and a significant contributor to re-ionizing the early universe. High-energy photons emitted by XRBs dominate the X-ray radiation field over active galactic nuclei \citep[AGN;][]{Jeon+22} at $z\gtrsim$6-8 \citep{Fragos+13a,Fragos+13b,Lehmer+22,Garofali+24,Zhang+24}, and the energy output of a single BH-HMXB can be orders of magnitude larger than core collapse supernovae \citep{Mirabel+11,Justham+12,Power+13}.

The specific frequency and total energy production of XRBs depends sensitively on their formation timescales, lifetimes, dynamic ranges, and duty cycles (DCs, the fraction of time when the X-rays are bright during the XRB phase of the binary system). Binary evolution models make predictions of these quantities, which can be constrained by resolved stellar population analysis (which places constraints on the ages of the optical counterpart to the X-ray source) and synoptic X-ray observations (for DC constraints). Thanks to its still-unrivaled spatial resolution and low background, observations from the \Chandra\ X-ray Observatory can identify XRBs down to low X-ray luminosities (\Lx$\sim$10$^{36}$ \lum) in the Local Volume (out to a few Mpc), where environmental conditions and even optical counterparts can be identified with complementary multiwavelength observations \citep[e.g.,][]{Antoniou+10,Antoniou+16,Binder+15,Binder+24,Lazzarini+21,Lazzarini+23}. While luminous XRBs (\Lx $\gtrsim10^{38}$ \lum) are generally powered by Roche lobe overflow, lower luminosity ($\sim$10$^{36}$ \lum) XRBs may accrete directly from the stellar wind of their companions \citep{Misra+23}, and the number and luminosity of XRBs formed depends on the structure of the donor star’s stellar wind \citep{Oskinova+11,Oskinova+12,MartinezNunez+17}. The most numerous class of these lower-\Lx\ XRBs are those with Be donor stars (BeXRBs), particularly in lower-metallicity environments \citep{Haberl+00,Liu+06,Antoniou+10,Antoniou+19}.

IC 10 is a nearby \citep[770 kpc;][]{Sanna+09a,Sanna+08b} dwarf irregular galaxy and one of the most active starburst galaxies in the Local Group. It hosts a young stellar population \citep[$\lesssim$6 Myr;][]{Massey+03,Massey+07} and one of the highest known spatial densities of Wolf-Rayet stars \citep{Crowther+03}. The discovery of the BH + WR binary IC 10 X-1 \citep{Bauer+04,Prestwich+07,Silverman+08} and the NS + potential luminous blue variable and supergiant fast X-ray transient (SFXT) IC 10 X-2 \citep{Laycock+14,Kwan+18,Alnaqbi+25} have motivated numerous observations with \Chandra. Its XRB population is believed to be dominated by HMXBs with supergiant donors \citep[sgHMXBs;][]{Laycock+17bsg}, rather than the more commonly observed SFXTs and XRBs with Be donors (BeXRBs) in the Milky Way \citep{Krivonos+12,Ducci+14} and Magellanic Clouds \citep{Antoniou+10}. Despite its close proximity, identifying optical counterparts to XRB candidates is difficult in IC~10 as it lies at a low Galactic latitude. However, most XRBs are highly variable X-ray emitters, and different subclasses of HMXBs tend to exhibit measurable differences in the properties of short- and long-term variability \citep{Sidoli+18}. X-ray variability is an important first-order indicator of the accretion mechanism powering an XRB, but the ubiquitous presence of XRBs that spend some fraction of time in a low-\Lx\ ``off'' state makes it difficult to empirically determine the total number of XRBs present in a galaxy, especially in a single ``snapshot'' observation. 

\citet[][hereafter L17]{Laycock+17var} presented a first look at the variability properties of X-ray sources in the IC 10 field as covered by Chandra ACIS-S3 in 9 visits spanning 2002-2010. Since then, we obtained an additional \Chandra\ observation of IC 10, which extended the observing baseline from $\sim$7 years to $\sim$19 years and allowed us to better constrain both the DCs and dynamic ranges (DRs; the ratio of the maximum ``outburst'' \Lx\ to the quiescent \Lx) of XRB candidates. Furthermore, {\em Gaia} observations can now help to separate foreground Galactic sources from those intrinsic to IC~10, significantly extending and improving on the positional counterpart identifications of L17b which were based on the \citet{Massey+07} catalog of ground-based photometry. In Section~\ref{sec:data} we describe the observations utilized in this work, data reduction procedures, and the cross-matching X-ray sources to {\em Gaia} counterparts. In Section~\ref{sec:variability} we discuss the short- and long-term variability properties observed for IC~10 sources. We present the XLF of IC~10 XRBs in Section~\ref{sec:xlf} and discuss the relationship between the observed XRB population, the star formation history (SFH) and metallicity evolution, and current recent star formation rate (SFR) in IC~10. We conclude with a discussion of our findings in Section~\ref{sec:discussion}. For consistency with previous SFH analyses, we assume throughout this work a distance to IC~10 of 770 kpc \citep{Sanna+08b,Sanna+09a} and adjust inferred X-ray source properties (e.g., \Lx) from the literature to this distance as needed. All uncertainties correspond to the 90\% confidence interval unless otherwise stated.

\section{Observations and Data Reduction}\label{sec:data}
We retrieved 11 publicly-available \Chandra/ACIS-S observations of IC~10, which are summarized in Table~\ref{tab:observation_log}\footnote{\bf See also \url{https://doi.org/10.25574/cdc.428}}. Ten of these observations (through the year 2010) were included in L17. Since then, two additional observations of IC~10 have been obtained: a deep ($\sim$150 ks, ObsID 15803) ACIS-I exposure and a $\sim$30 ks ACIS-S exposure (ObsID 26188). ObsID 15803 was taken in 1/8-subarray mode for the purpose of studying IC~10 X-1, and thus nearly all other X-ray point sources that would have otherwise been available to the ACIS-I detector were excluded from the field of view. Given the numerous detailed studies of IC~10 X-1 \citep{,Laycock+15a,Laycock+15b,Steiner+16,Bhattacharya+23a,Bhattacharya+23b,Wang+24}, we do not include that source in our study of X-ray variability and therefore do not use ObsID 15803 in the present work (IC~10 X-1 is, however, included in our analysis of the XLF; see Section~\ref{sec:xlf}). The introduction of ObsID 26188, which is approximately twice as deep as the \Chandra\ monitoring campaign of IC~10 that spanned 2009-2010, aids in refining X-ray positions and extending the baseline of observations from $\sim$7 years considered in L17 to $\sim$19 years in the current study. In this study we incorporate the entire Chandra ACIS field of view, in contrast to L17a,b who restricted their analysis to the single ACIS-S3 chip (back illuminated CCD), although S3 field of view makes up the majority of the 2MASS $K_S$ isophotal area.  A joint \Chandra\ and {\em James Webb Space Telescope} (JWST) observation of IC~10 is in progress, and will enable an unprecedented look at the environments in which these HMXBs reside.

\begin{table}
\centering
\caption{\Chandra\ ACIS-S Observation Log}\label{tab:observation_log}
\setlength{\tabcolsep}{3pt}
    \begin{tabular}{cccccc}
    \hline \hline
            & R.A. & Decl.              &   &  & Exp. Time    \\ \cline{2-3}
    ObsID   & \multicolumn{2}{c}{(J2000)} & Date  & MJD & (ks)    \\
    (1) & (2) & (3) & (4) & (5) & (6) \\
    \hline
    3953    & 00:20:25  & 59:16:55  & 12 Mar 2003   & 52710 & 28.9    \\
    7082    & 00:20:04  & 59:16:45  & 02 Nov 2006   & 54041 & 40.1    \\
    8458    & 00:20:04  & 59:16:45  & 04 Nov 2006   & 54044 & 40.5   \\
    11080   & 00:20:17  & 59:17:56  & 05 Nov 2009   & 55140 & 14.6    \\
    11081   & 00:20:19  & 59:18:02  & 25 Dec 2009   & 55190 & 8.1    \\
    11082   & 00:20:23  & 59:17:10  & 11 Feb 2010   & 55238 & 14.7    \\
    11083   & 00:20:34  & 59:19:01  & 04 Apr 2010   & 55290 & 14.7    \\
    11084   & 00:20:35  & 59:20:16  & 21 May 2010   & 55337 & 14.2    \\
    11085   & 00:20:11  & 59:19:13  & 20 Jul 2010   & 55397 & 14.5    \\
    11086   & 00:20:15  & 59:18:11  & 05 Sep 2010   & 55444 & 14.7    \\
    26188   & 00:20:29  & 59:16:52  & 07 Jan 2022   & 59586 & 29.7    \\
    \hline \hline
    \end{tabular}
\end{table}

All observations were uniformly reduced using the \Chandra\ Interactive Analysis of Observations \citep[CIAO;][]{Fruscione+06} software version 4.16 and calibration database (CALDB) version 4.11.5 following standard reduction procedures. Data were reprocessed from \texttt{evt1} using the CIAO task \texttt{chandra\_repro} and we used the point source detection tool \texttt{wavdetect} to identify a preliminary list of point sources in each individual exposure. The major and minor axes of each preliminary source's error circle were increased by a factor of 5 to account for residual potential source photons from the wings of the point spread function, and the X-ray sources were masked so that a background light curve for the full observation could be extracted. These background light curves were inspected for background flares, and good time intervals (GTIs) were generated using the \texttt{lc\_clean} algorithm. We filtered the reprocessed \texttt{evt2} files on these GTIs and restricted the energy range to 0.5-7 keV.

We used \texttt{wcs\_match} to align all observations to ObsID 8458, which had the deepest exposure time, to update the absolute astrometry. Images were reprojected and combined using the tasks \texttt{reproject\_obs} and \texttt{flux\_obs}. We finally re-ran \texttt{wavdetect} on the cleaned, merged image using a significance threshold of 10$^{-6}$ and smoothing scales of ``1 2 4 8 16''. A total of 156 sources were detected in the merged \Chandra\ image.

\subsection{Excluding Foreground and Background Sources}
\begin{figure}
    \centering
    \includegraphics[width=1\linewidth,clip=true,trim=3cm 8cm 2cm 7.5cm]{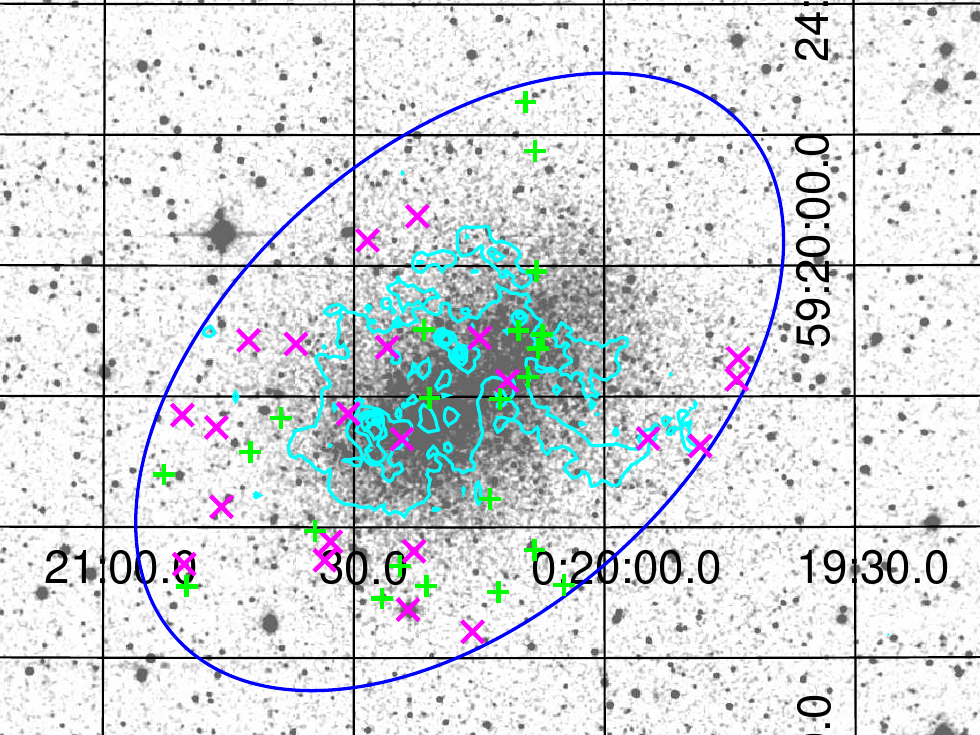}
    \caption{The 2MASS $K_S$ image of IC~10, with the isophotal radius is shown in blue. Foreground stars identified by {\em Gaia} are shown in magenta (``x'' markers) and X-ray sources likely intrinsic to IC~10 are shown in green (``+'' markers). H$\alpha$ contours are shown in cyan.}
    \label{fig:merged_with_sources}
\end{figure}

To identify HMXBs residing within IC~10, we first excluded all sources beyond the 2MASS Large Galaxy Atlas $K_s$ ``total'' magnitude isophotal diameter \citep{Jarrett+03}, which has a major axis of $\sim$0.195$^{\circ}$ (corresponding to $\sim$3 kpc at the distance of IC~10), an axis ratio of 0.61, and a position angle of 132$^{\circ}$ (i.e., the blue oval in Figure~\ref{fig:merged_with_sources}). We then kept only the 46 X-ray sources that were detected at $\geq$5$\sigma$ in the merged image. 

We next cross-matched our X-ray source list with {\em Gaia} \citep{GaiaDR3} detections. \citet{He+23} found that some IC~10 stars were detected by {\em Gaia}, with a median parallax of 0.023 mas. We therefore removed 20 X-ray sources that are coincident (within a 2$^{\prime\prime}$) with a {\em Gaia} star with a parallax $\geq$0.08 mas in magnitude (corresponding to a distance of $\sim$12 kpc) and measured at $\geq$3$\sigma$ significance, as these are most likely foreground Galactic sources. We found 15 X-ray sources that were coincident with {\em Gaia} stars and have firm parallax measurements that place them within $\sim$1.5 kpc of the Sun. An additional five sources had less-secure parallax measurements (the parallax $p$ divided by the parallax uncertainty, $p/\sigma_p$, was $<$3) but their high proper motions were consistent with nearby Galactic stars; we exclude these stars from our HMXB candidates list. Table~\ref{tab:galactic_stars} summarizes the stellar parameters derived from {\em Gaia} for X-ray sources with an optical {\em Gaia} counterpart. Columns 1-2 provide the R.A. and Decl. coordinates (J2000) for each X-ray source as measured from the \Chandra\ observations. The third column provides the source ID number given in L17. The fourth column provides the ratio of the {\em Gaia} parallax ($p$) to the parallax uncertainty ($\sigma_p$), both of which are measured in units of mas, and column 5 provides the distance to the {\em Gaia} counterpart. Columns 6-7 provide the proper motions of the {\em Gaia} counterpart (in both R.A., $\alpha$, and Decl., $\delta$). Columns 8-11 provide the apparent $G$ magnitude, the $G$-band extinction $A_G$, the color excess $E_{BP-RP}$, and the source color $BP-RP$. Columns 12-14 provide the photospheric effective temperature $T_{\rm eff}$, log$g$, and [M/H] derived from Gaia spectroscopy\footnote{See Section 6 of \url{https://gea.esac.esa.int/archive/documentation/GDR3/index.html}.} of the optical counterparts.

To determine the probability of a chance coincidence between one of these {\em Gaia} stars and an X-ray source, we randomly distributed an equal number of simulated {\em Gaia} stars and X-ray sources as in our observed samples and measured the frequency with which a simulated {\em Gaia} star overlapped with a simulated X-ray source, using the same 2$^{\prime\prime}$ matching radius as with the observed data. We performed 1000 iterations of randomly distributing simulated sources and found the probability of a chance coincidence was $\sim$6.4\%. Thus, it is possible that $\sim$1-2 X-ray sources that matched to a Galactic star according to our criteria are chance coincidences.

\begin{table*}
\centering
\caption{{\em Gaia} Properties of Likely Galactic Foreground Stars}\label{tab:galactic_stars}
    \scriptsize
    \setlength{\tabcolsep}{2pt}
    \begin{tabular}{cccccccccccccc}
    \hline \hline
    RA  & Decl.                 & L17   & parallax ($p/\sigma_p$)  & distance & $\mu_{\alpha}$    & $\mu_{\delta}$    & $G$   & $A_G$ & $E_{BP-RP}$ & $BP-RP$ & $T_{\rm eff}$  & log$g$ & [M/H] \\ \cline{1-2}
    \multicolumn{2}{c}{(J2000)} &  & (mas) & (pc) & (mas) & (mas) & (mag) & (mag) & (mag) & (mag) & (K) & [cm s$^{-2}$] & \\
    (1) & (2) & (3) & (4) & (5) & (6) & (7) & (8) & (9) & (10) & (11) & (12) & (13) & (14) \\
    \hline
    5.0658221 & 59.2400576   & 3    & 0.416$\pm$0.376 (1.11)    & \nodata             & -1.793 & -0.803 & 15.54  & \nodata                 & \nodata                & 2.13  & \nodata              & \nodata                & \nodata                   \\     
    5.0979852 & 59.2457214   & 5    & 3.049$\pm$0.013 (234.54)  & 325$^{+5}_{-4}$     &  8.724 & -6.895 &  9.98  & 0.56$^{+0.02}_{-0.03}$  & 0.30$\pm$0.01          & 0.86  & 6581$^{+42}_{-51}$   & 3.71$^{+0.01}_{-0.02}$ & -0.37$^{+0.02}_{-0.03}$    \\ 
    5.1108909	& 59.248712  & 7    & 0.085$\pm$0.042 (2.02)    & \nodata             & -1.706 & -1.016 & 16.23  & \nodata                 & \nodata                & 1.97  & \nodata              & \nodata                & \nodata                    \\  
    4.9781502 & 59.2893036   & 26   & 2.876$\pm$0.654 (4.40)    & \nodata             & 7.220  & -2.149 & 20.42  & \nodata                 & \nodata                & 2.24  & \nodata              & \nodata                & \nodata                    \\
    5.1937029 & 59.2921289   & 28   & 1.050$\pm$0.110 (9.55)    & 629$^{+25}_{-21}$   & -3.466 & -2.201 & 18.08  & 0.50$^{+0.04}_{-0.03}$  & 0.27$\pm$0.02          & 1.99  & 3752$^{+29}_{-23}$   & 4.42$^{+0.04}_{-0.02}$ & -1.34$^{+0.06}_{-0.04}$    \\ 
    5.0484231 & 59.3040421   & 32   & 0.810$\pm$0.265 (3.06)    & \nodata             & 3.334  &  0.315 & 19.44  & \nodata                 & \nodata                & 2.40  & \nodata              & \nodata                & \nodata                    \\
    5.1084374 & 59.3125939   & 39   & 1.169$\pm$0.100 (11.69)   & 502$^{+55}_{-43}$   & -2.084 & -3.737 & 17.95  & 0.77$^{+0.29}_{-0.05}$  & 0.43$^{+0.16}_{-0.03}$ & 2.16  & 3846$^{+124}_{-38}$  & 4.80$^{+0.05}_{-0.06}$ & -0.61$^{+0.17}_{-0.21}$    \\
    5.1777308 & 59.3143043   & 48   & 1.071$\pm$0.072 (14.88)   & 1073$^{+188}_{-83}$ & -2.380 &  0.640 & 17.42  & 0.63$\pm$0.02           & 0.34$\pm$0.01          & 1.82  & 4286$^{+19}_{-26}$   & 4.43$^{+0.06}_{-0.13}$ & -0.10$\pm$0.04             \\
    5.1387865 & 59.3595756   & 52   & 1.904$\pm$0.051 (37.33)   & 502$^{+8}_{-6}$     & 4.215  &  0.132 & 16.67  & 0.80$^{+0.02}_{-0.01}$  & 0.43$\pm$0.01          & 1.93  & 4308$^{+11}_{-7}$    & 4.67$\pm$0.01          & 0.03$\pm$0.03              \\ 
    5.1365133 & 59.2629980   & 61   & 2.470$\pm$0.311 (7.94)    & \nodata             & 0.789  & -3.451 & 19.54  & \nodata                 & \nodata                & 2.72  & \nodata              & \nodata                & \nodata                    \\
    5.0949905 & 59.2605903   & 64   & 0.730$\pm$0.021 (34.76)   & 1564$^{+144}_{-90}$ & -2.664 & -3.830 & 14.92  & 1.31$^{+0.02}_{-0.04}$  & 0.71$\pm$0.02          & 1.09  & 7224$^{+61}_{-102}$  & 4.13$^{+0.05}_{-0.07}$ & -0.61$^{+0.03}_{-0.04}$    \\
    5.1514249 & 59.3687520   & 85   & 2.388$\pm$0.182 (13.12)   & 326$^{+15}_{-12}$  & 6.239  & -2.758  & 18.81  & 0.83$\pm$0.05           & 0.57$\pm$0.03          & 3.03  & 3265$^{+19}_{-17}$   & 4.89$\pm$0.02          & 0.05$\pm$0.03              \\
    5.1540232 & 59.3134252   & 86   & 0.783$\pm$0.123 (6.37)    & 685$^{+82}_{-57}$  & -0.240 & -3.470  & 18.34  & 0.39$^{+0.08}_{-0.06}$  & 0.21$^{+0.05}_{-0.03}$ & 1.66  & 4146$^{+74}_{-47}$   & 4.96$^{+0.04}_{-0.05}$ & -1.49$^{+0.10}_{-0.08}$    \\
    5.0381315 & 59.3925278   & 87   & 0.305$\pm$0.062 (4.92)    & \nodata            & -1.105 &  0.287  & 17.12  & \nodata                 & \nodata                & 1.50  & \nodata              & \nodata               & \nodata                    \\ 
    5.1182849 & 59.3398220   & 90   & 0.521$\pm$0.535 (0.97)    & \nodata            & -2.506 & -1.893  & 20.07  & \nodata                 & \nodata                & 2.48  & \nodata              & \nodata               & \nodata                     \\ 
    5.0427644 & 59.317039    & 106 & -0.126$\pm$0.670 (0.19)    & \nodata            & 0.634  & -1.976  & 20.52  & \nodata                 & \nodata                & 2.29  & \nodata              & \nodata               & \nodata                     \\
    5.2096170 & 59.2572167   & 107     & 0.403$\pm$0.140 (2.88) & 242$\pm$7          & 31.540 &  1.232  & 18.48  & 0.77$\pm$0.04           & 0.52$^{+0.03}_{-0.02}$ & 3.07  & 3250$^{+10}_{-9}$    & 4.96$^{+0.01}_{-0.02}$ & -0.11$\pm$0.03            \\ 
    5.1394606 & 59.2582661   & \nodata & 0.889$\pm$0.250 (3.56) & \nodata            & 8.223 & -4.106   & 19.23  & \nodata                 & \nodata                & 2.16  & \nodata              & \nodata               & \nodata                    \\
    5.1910708 & 59.2718642   & \nodata & 0.954$\pm$0.281 (3.40) & \nodata            & -2.800 & -1.833  & 19.35  & \nodata                 & \nodata                & 2.19  & \nodata              & \nodata               & \nodata                    \\ 
    5.0934138 & 59.3459836   & \nodata & 0.798$\pm$0.153 (5.22) & 706$^{+137}_{-66}$ &  6.627 & -5.640  & 18.61  & 0.40$^{+0.05}_{-0.04}$  & 0.22$\pm$0.03          & 2.03  & 3697$^{+74}_{-44}$    & 4.76$^{+0.07}_{-0.09}$ & -0.61$^{+0.19}_{-0.16}$   \\

    \hline \hline
    \end{tabular}
\end{table*}

Of the remaining 26 X-ray sources without a {\em Gaia} counterpart, one was classified by L17 as a suspected Galactic source due to its low absorption and extremely soft X-ray flux. We next searched for counterparts from the Zwicky Transient Factory\footnote{\url{https://www.ztf.caltech.edu/}} \citep[ZTF;][]{ZTF1,ZTF2} and identified three X-ray sources with well-sampled optical counterpart light curves (we note that all but one of the likely Galactic sources listed in Table~\ref{tab:galactic_stars} had a well-sampled ZTF light curve). One of these is the SFXT IC~10 X-2 \citep{Kwan+18}, which we retain in our sample, while the other two sources exhibit colors, magnitudes, and optical flaring consistent with Galactic main sequence M dwarfs and are removed from our candidate HMXBs list. 

Our final candidate HMXBs list contains 23 sources, shown in Figure~\ref{fig:merged_with_sources} and summarized in Table~\ref{tab:hmxbs}. This number is broadly consistent with the predicted number of HMXBs above $\sim$a few$\times10^{35}$ \lum\ from the XLFs of galaxies of similar metallicities to IC~10 \citep{Lehmer+21}, although there is considerable uncertainty in the recent SFR of IC~10 (see Section~\ref{sec:xlf}). We use WebPIMMS\footnote{See \url{https://heasarc.gsfc.nasa.gov/cgi-bin/Tools/w3pimms/w3pimms.pl}} to derive a conversion factor between count rates and fluxes for IC~10 sources detected with the \Chandra\ ACIS-S instrument. We assume a power law with $\Gamma=1.7$ (appropriate for HMXBs, as inferred from spectral modeling results in L17) and a foreground absorbing column of $5.02\times10^{21}$ cm$^{-2}$ \citep{HI4PI}. Assuming this spectral model, 1 ct s$^{-1}$ equates to an unabsorbed flux of 1.98$\times10^{-11}$ \flux. The detection limit of the merged observation is $\sim$1.7$\times10^{-15}$ \flux, which corresponds to a luminosity of $\sim$1.2$\times10^{35}$ \lum\ at the distance of IC~10. The individual exposures have shallower detection limits on the order of several$\times10^{35}$ \lum. Within the 2MASS $K_s$ $D_{25}$ isophotal diameter of IC~10, the \citet{Cappelluti+09} AGN log$N$-log$S$ distribution predicts $\sim$5-7 X-ray emitting AGN are expected. Table~\ref{tab:hmxbs} also contains information on the X-ray variability properties of each HMXB candidate, which is discussed in Section~\ref{sec:variability}.

\begin{table*}
\scriptsize
\centering
\caption{X-ray Properties of HMXB Candidates in IC~10}\label{tab:hmxbs}
\setlength{\tabcolsep}{2pt}
    \begin{tabular}{ccccccccccccc}
    \hline \hline
    RA  & Decl.               & L17 & \# of Det- & \# Obs Out- & Signif-  & Net Counts & \multicolumn{2}{c}{Short-Term Constant CR$^*$}  && \multicolumn{2}{c}{Long-Term Variability} & max \Lx/$10^{35}$ \\ \cline{1-2} \cline{8-9} \cline{11-12}
    \multicolumn{2}{c}{(J2000)} & ID & ections & side FOV & icance ($\sigma$)    & (0.5-7 keV) & Rejected    & Not Rejected       && DR      & DC     & (\lum)    \\
    (1) & (2) & (3) & (4) & (5) & (6) & (7) & (8) & (9) && (10) & (11) & (12)  \\
    \hline
    5.0528153	& 59.250361   & 8 & 11 & 0 & 141.3    & 1043$\pm$33 & 0 & 11 && 1.9      & $>$55            & 36.3$^{+5.9}_{-4.8}$  \\  
    5.0887259	& 59.251876   & 9 & 2  & 0  & 10.0     & 67$\pm$10 & 0 &  2  && 31.2     & 15$^{+50}_{-15}$ & 3.3$^{+1.8}_{-0.2}$  \\  
    5.0202434	& 59.252087   & 10 & 6  & 0  & 18.7     & 137$\pm$13 & 0 &  6  && 104.3    & 35$^{+30}_{-15}$ & 7.1$^{+2.6}_{-1.0}$  \\   
    5.1020043	& 59.257017   & 12 & 11 & 0 & 162.8    & 1351$\pm$37 & 0 & 11  && 2.5      & $>$55            & 49.3$^{+8.0}_{-6.6}$  \\  
    5.0347953	& 59.261108   & 14 & 8  & 0  & 25.7     & 183$\pm$15 & 0 & 7  && 174.9    & 70$^{+10}_{-35}$ & 12.1$^{+4.0}_{-2.0}$  \\ 
    5.0569970	& 59.274141   & 18 & 8  & 0  & 79.7     & 459$\pm$22 & 0 & 8  && 202.8    & 65$^{+20}_{-25}$ & 13.9$^{+4.5}_{-2.4}$  \\  
    5.1010337   & 59.289176   & 25 & 1 & 0 &  5.1 & 22$\pm$6 & 0 & 1 && 46.5 & $<$45 & 2.0$^{+0.5}_{-0.2}$ \\
    5.0380233	& 59.305163   & 29 & 5  & 0  & 21.9     & 98$\pm$10 & 0 & 5  && 97.6     & 50$^{+10}_{-20}$ & 5.1$^{+1.9}_{-0.7}$  \\   
    5.1278952   & 59.295631 & 30 & 1 & 0 & 9.8 & 64$\pm$10 & 0 &  1 && 108.2 & 15$^{+35}_{-15}$ & 6.0$^{+2.4}_{-0.9}$ \\
    4.9338027   & 59.304295 & 33 & 9 & 0 & 14.9 & 162$\pm$16 & 0 & 9 && 73.6 & 85$^{+10}_{-20}$ & 11.7$^{+4.4}_{-1.6}$ \\
    5.0330476	& 59.312431 & 38 & 1  & 0  & 6.3      & 23$\pm$5& 0 & 1   && 44.4     & 10$^{+50}_{-5}$  & 2.2$^{+0.5}_{-0.2}$  \\  
    5.0623012   & 59.315010 & 40 & 8 & 0 & 51.1 & 276$\pm$17 & 0 & 8 & & 171.8 & 65$^{+10}_{-20}$ & 8.6$^{+4.0}_{-1.3}$ \\
    5.0307852	& 59.315977   & 41 & 4  & 0  & 26.0     & 179$\pm$15 & 0 & 4  && 222.3    & 20$^{+30}_{-15}$ & 8.3$^{+1.5}_{-0.9}$  \\  
    5.0897976	& 59.317258   & 42 & 7  & 0  & 34.6     & 165$\pm$13 & 1  & 6  && 183.3    & 50$^{+25}_{-20}$ & 8.7$^{+3.0}_{-1.4}$  \\   
    5.0345288	& 59.362871   & 43 & 9  & 1  & 19.8     & 256$\pm$20 & 0 & 9  && 349.9    & 75$^{+15}_{-30}$ & 16.4$^{+5.1}_{-2.9}$  \\  
    5.1612495	& 59.294670   & 45 & 6  & 0  & 21.1     & 141$\pm$13 & 0 & 6  && 129.2    & 60$^{+15}_{-35}$ & 7.2$^{+2.6}_{-1.1}$  \\  
    5.0871851   & 59.299774   & 46 & 2 & 0 & 65.1 & 338$\pm$19 & 0 & 2 && 1309.7 & 25$^{+30}_{-20}$ & 68.4$^{+13.5}_{-11.0}$ \\
    5.0391930	& 59.375325   & 57 & 2  & 1  & 8.1      & 84$\pm$13 & 0 & 2  && 86.4     & 10$^{+35}_{-5}$ & 4.8$^{+2.0}_{-0.6}$  \\   
    5.0519087	& 59.299533   & 58 & 5  & 0  & 28.6     & 125$\pm$12 & 0 & 5  && 100.2    & 30$^{+15}_{-10}$ & 5.5$^{+1.5}_{-0.6}$  \\   
    5.0337290	& 59.332104   & 79 & 1  & 1  & 10.7     & 71$\pm$10 & 0 & 1  && 119.8    & 10$^{+45}_{-5}$ & 5.6$^{+2.3}_{-0.8}$  \\    
    5.2196343	& 59.280225   & 82 & 3  & 1  & 13.6     & 101$\pm$12 & 0 & 3  && 109.7    & 35$^{+60}_{-30}$ & 10.0$^{+4.1}_{-1.2}$  \\  
    5.2082830	& 59.251743   & 95 & 1  & 4  & 6.5      & 42$\pm$9 & 0 & 1  && 76.2     & 15$^{+50}_{-10}$ & 6.3$^{+2.4}_{-0.8}$  \\  
    5.1766139	& 59.286000   & 108 & 1  & 0  & 10.7     & 59$\pm$9 & 0 & 1  && 182.3    & 15$^{+40}_{-10}$ & 10.1$^{+2.4}_{-1.3}$ \\  
    \hline \hline
    \multicolumn{13}{l}{$^*$The number of observations in which the null hypothesis of a constant count rate (CR) was rejected or not rejected at the 1\%} \\
    \multicolumn{13}{l}{significance level.}
    \end{tabular}
\end{table*}

\section{X-ray Variability}\label{sec:variability}
The variability of accreting compact objects can be studied in two ways: inter-observation variability, which we refer to as ``long-term'' variability in which the X-ray emission significantly changes over timescales of weeks, months, or years, and intra-observation variability that occurs on timescales of hours or less (which we refer to as ``rapid'' variability). Understanding X-ray variability across different timescales reveals information about the mass transfer mechanism, accretion physics, and potentially the wind structure between donor star and compact object when viewed on rapid timescales, while longer-term monitoring yields information on burst-like behavior and possible state transitions within the HMXB. We discuss both the long-term and rapid variability characteristics of the IC~10 HMXB candidate sample below.

\subsection{Long-Term Variability}
We use the coordinates for all HMXB candidates that were robustly detected ($\geq$5$\sigma$) in the merged \Chandra\ image of IC~10 as input for the CIAO task \texttt{srcflux}, using the ``arfcoor'' tool to generate a model of the point spread function (PSF) at each source location and the spectral model described in the previous section. The \texttt{srcflux} task then returns, among other quantities, the net count rate, absorbed and unabsorbed fluxes, source significance, and determines if the source was within the detector field of view and if the count rates and fluxes reflect upper limits or firm detections.

\begin{figure}
    \centering
    \includegraphics[width=1\linewidth,clip=true,trim=0.5cm 0.25cm 0cm 0cm]{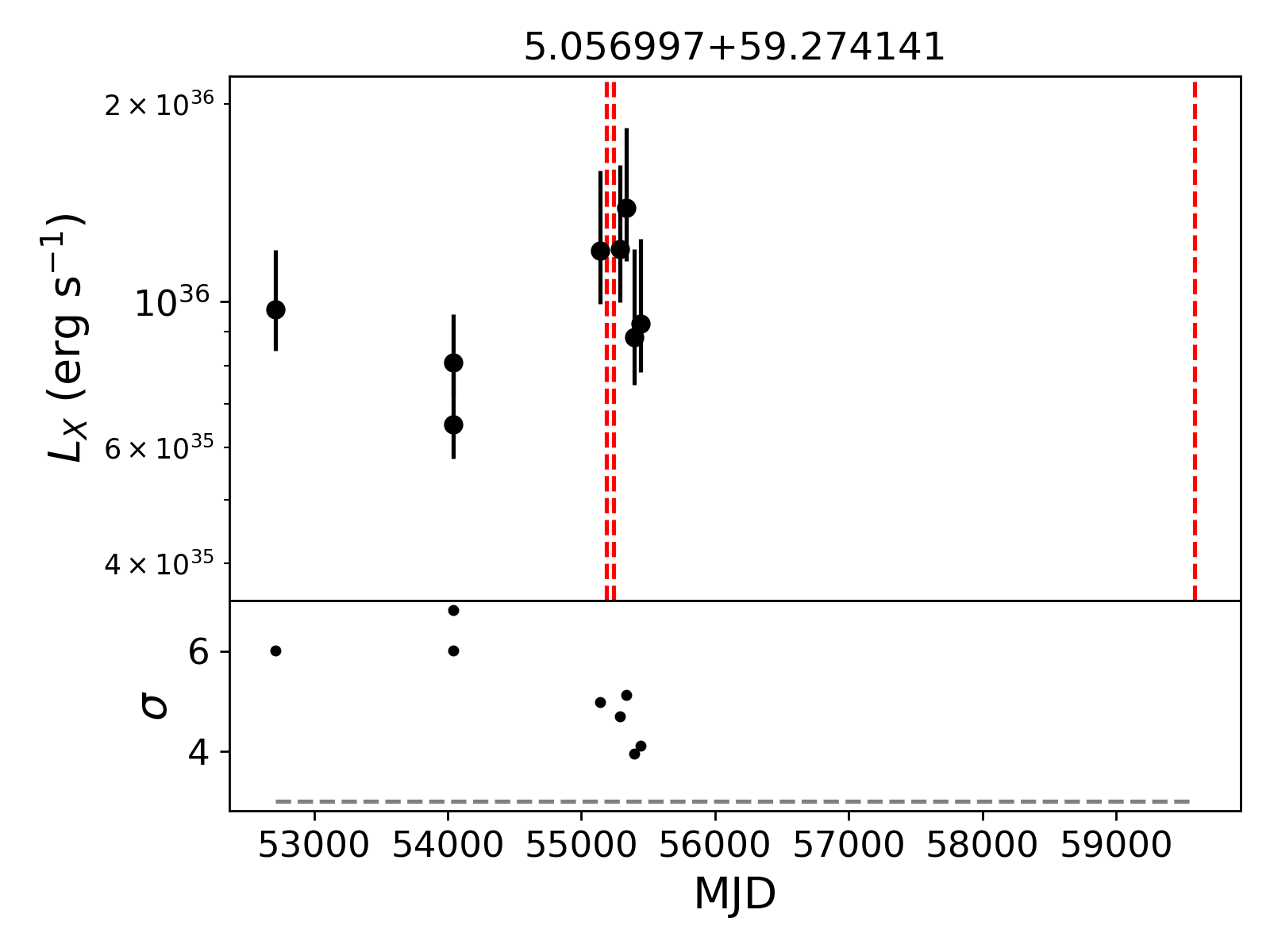}
    \caption{The long-term light curve of HMXB candidate located at R.A.=5.0569970$^{\circ}$ and Decl.=+59.274141$^{\circ}$. {\it Top}: \Lx\ as a function of time (black circles with error bars). Red, dashed vertical lines show the dates of observations in which the source was not detected. {\it Bottom}: the significance at which the source was detected in each observation.}
    \label{fig:lc_long_ex}
\end{figure}

Determining a source's DC requires knowledge of when a source was observed to be ``on'' (in a detected high-emission state) and when it was ``off'' (that is, not detected). Figure~\ref{fig:lc_long_ex} shows an example long-term light curve for the source detected at R.A.=5.0569970$^{\circ}$ and Decl.=+59.274141$^{\circ}$. Both $\geq$3$\sigma$ detections (black points) and non-detected epochs (indicated with red dashed lines) are shown, and we show both the \Lx\ and the detection significance as functions of time. In this case, the source location was within the field of view of all eleven \Chandra\ pointings, and it was confidently detected (i.e., the source was ``on'') in nine observations but fell below the detection limit (the source was ``off'') in two observations. We measured the 0.5-7 keV DR as the ratio of the observed maximum to minimum \Lx\ for sources that were detected in all observations in which they were contained within the field of view. For sources that were undetected in one or more observations, we calculated the lower limit on the DR using the lowest detection limit of the observations in which the source was undetected.

To constrain the X-ray DCs of sources in the IC~10 sample, we emulated the approach utilized by \citet{Kyer+24} in their study of the variability properties of HMXBs in M33; a summary of this approach is described here, and the reader is referred to \citet{Kyer+24} for further details. We generated synthetic light curves assuming a grid of DCs and timescales over which each simulated source was ``on.'' The DC grid ranged from 5\% to 100\% in intervals of 5\%, and the timescales we sampled included a 1-10 day range (at 1 day intervals) and 10-200 days (at 10 day intervals). For each combination of DC and timescale, we generated 500 synthetic light curves (for a total of 290,000 light curves). A simulated light curve is designated as being in the ``on'' state for a number of days equal to the timescale of variability multiplied by the DC. While all the days that a source is ``on'' are sequential, the first day which the source turns ``on'' (i.e., the phase) in each light curve is random. We divide the total observation baseline (6876 days) by the variability timescale, rounding up, and then adding an extra cycle. This allows us to randomize the start date of the first ``on'' epoch for each simulated source. 

\begin{figure}
    \centering
    \includegraphics[width=1\linewidth,clip=true,trim=0cm 0.5cm 0cm 2cm]{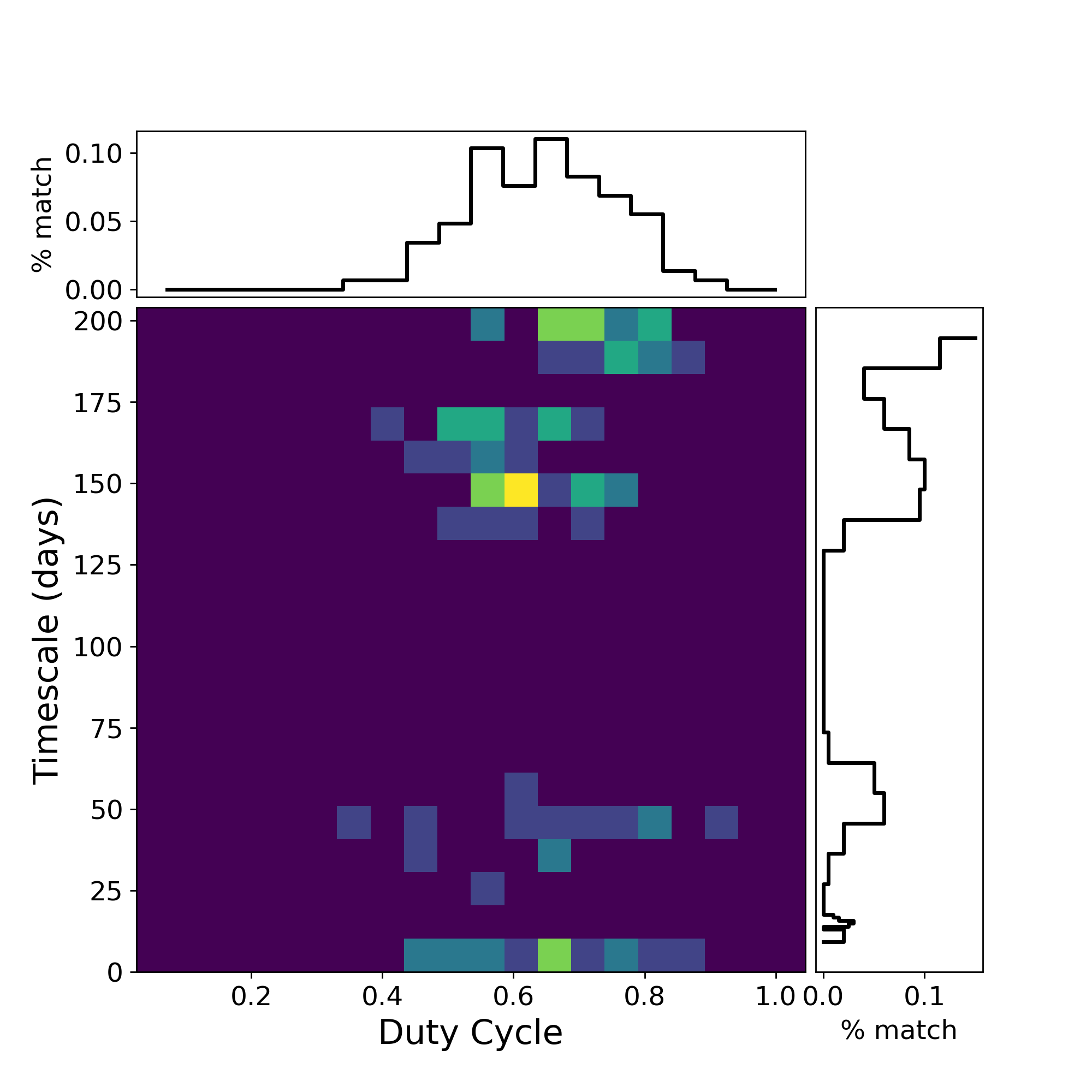}
    \caption{The match percentage between observed and simulated light curves for HMXB candidate located at R.A.=5.0569970$^{\circ}$ and Decl.=+59.274141$^{\circ}$. {\it Main window}: dark blue indicates 0\% matches, and lighter/warmer colors show increasing match frequencies. Only the 10-200 day timescales are shown for clarity. {\it Top}: the match frequency as a function of DC. {\it Right}: the match frequency as a function of variability timescale.}
    \label{fig:constraint_table}
\end{figure}

The synthetic light curves were then sampled at the same time intervals as the observations. We compared the epochs in which the IC~10 sources were observed to be ``on'' and ``off'' with the simulated light curves and recorded the frequency with which the observed and synthetic light curves matched for each source (i.e., the source activity was the same on all days of observations between the observed and simulated sources). Figure~\ref{fig:constraint_table} shows the fraction of matching synthetic light curves for the same source as in Figure~\ref{fig:lc_long_ex} as a function of DC and variability timescale. In general, the timescale of variability is not well constrained by the data, and this method is unable to reliably constrain DCs $\lesssim$5\%. We report the DC with the highest match frequency between observed and simulated observations as the ``best'' DC. The upper and lower limits on the DC are determined by the maximum and minimum DCs that yielded at least one match between observed and simulated light curves.

\begin{figure}
    \centering
    \includegraphics[width=1\linewidth]{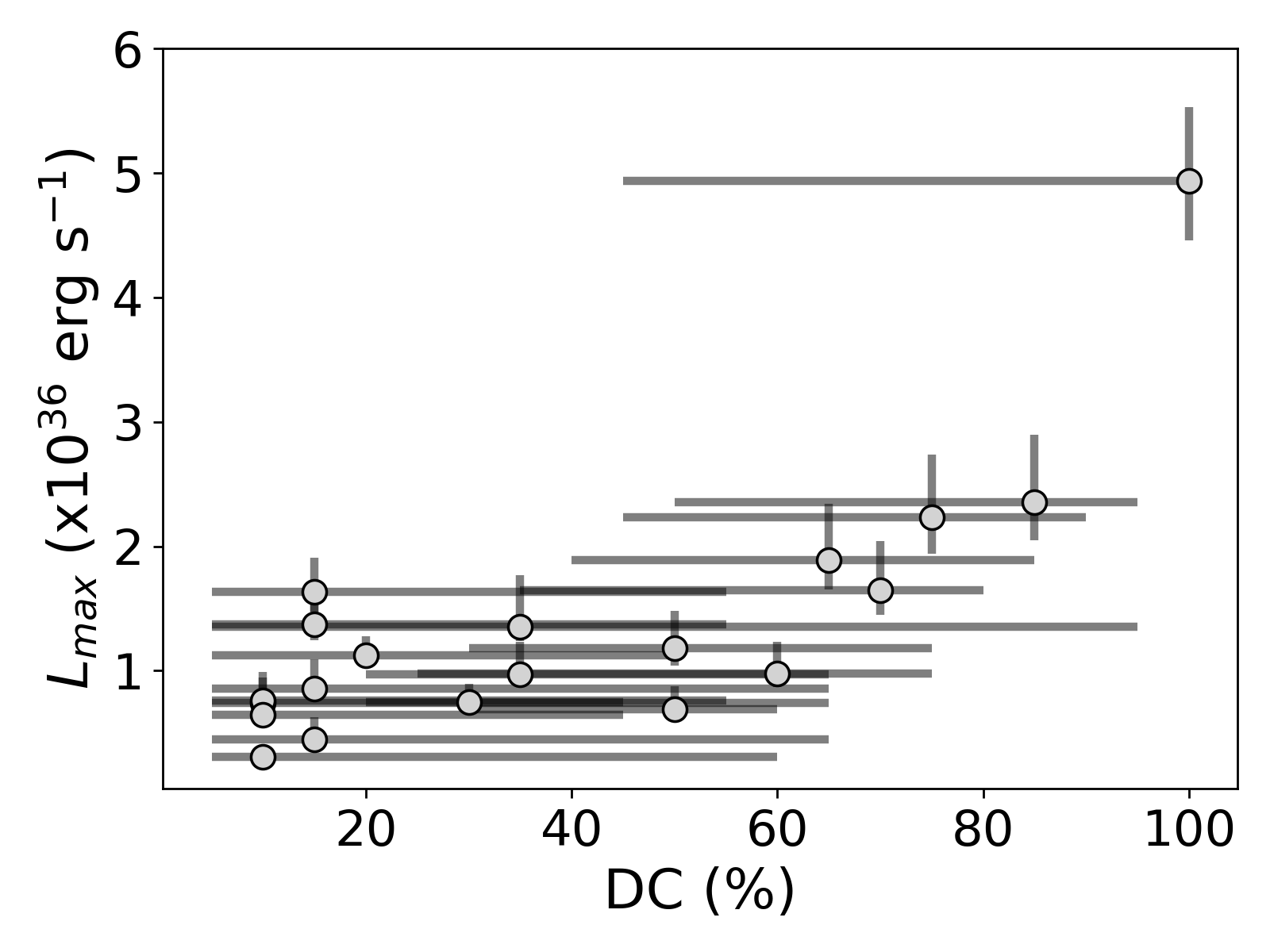}
    \caption{The maximum observed \Lx\ of each IC~10 HMXB candidate as a function of DC. The two sources with the highest peak \Lx\ are the only two with DCs consistent with 100\% (persistently bright).}
    \label{fig:Lmax_vs_DC}
\end{figure}

\citet{Kyer+24} found that M33 with lower peak count rates had a systematically higher fraction of ``off'' epochs than sources with higher peak count rates, indicating that faint sources may be more likely to turn off than bright sources. We investigated this for the IC~10 sources by comparing the inferred DC to the maximum observed \Lx\ (which we denote $L_{\rm max}$) for each source. As shown in Figure~\ref{fig:Lmax_vs_DC}, the brightest source in our IC~10 sample has a DC consistent with 100\%, and there is a correlation between lower $L_{\rm max}$ and lower DCs (albeit with large DC uncertainties). While count rate fluctuations in sources near the M33 survey limit could have exaggerated the trend in M33, we find evidence of sources with high $L_{\rm max}$ (all of which are well above the detection limits of the \Chandra\ observations of IC~10) having higher DCs in IC~10. Our analysis supports the idea that brighter sources ($\gtrsim4\times10^{36}$ \lum) tend to be more frequently in the bright state than fainter sources.

\subsection{Rapid Variability}
We next quantified the short-term, intra-observation variability exhibited by each source in all of the observations in which it was detected at $\geq3\sigma$ significance. We used the CIAO task \texttt{dmextract} to extract background-subtracted light curves for all sources that were detected at $\geq3\sigma$ significance (as determined by the \texttt{srcflux} task) in each individual \Chandra\ observation. While Kolmogorov-Smirnov (KS) test is commonly employed to assess variability in astrophysical sources, we use the Anderson-Darling test, which is more sensitive to short duration variations in X-ray light curves \citep{Feigelson+22}. 

We use the \texttt{scipy} routine \texttt{anderson\_ksamp} to compare the cumulative photon count distributions from the observed (unbinned) light curves to the photon count distributed expected for a perfectly constant count rate source. We use the returned $p$-values to determine that, for the majority sources, the null hypothesis of a constant count rate (CR) cannot be rejected at the 1\% significance level. We did not find a significant increase in null hypothesis rejections if the significance threshold was relaxed to 5\%; thus, very few sources show strong evidence for rapid CR variability in the present data set.

\begin{figure}
    \centering
    \includegraphics[width=1\linewidth]{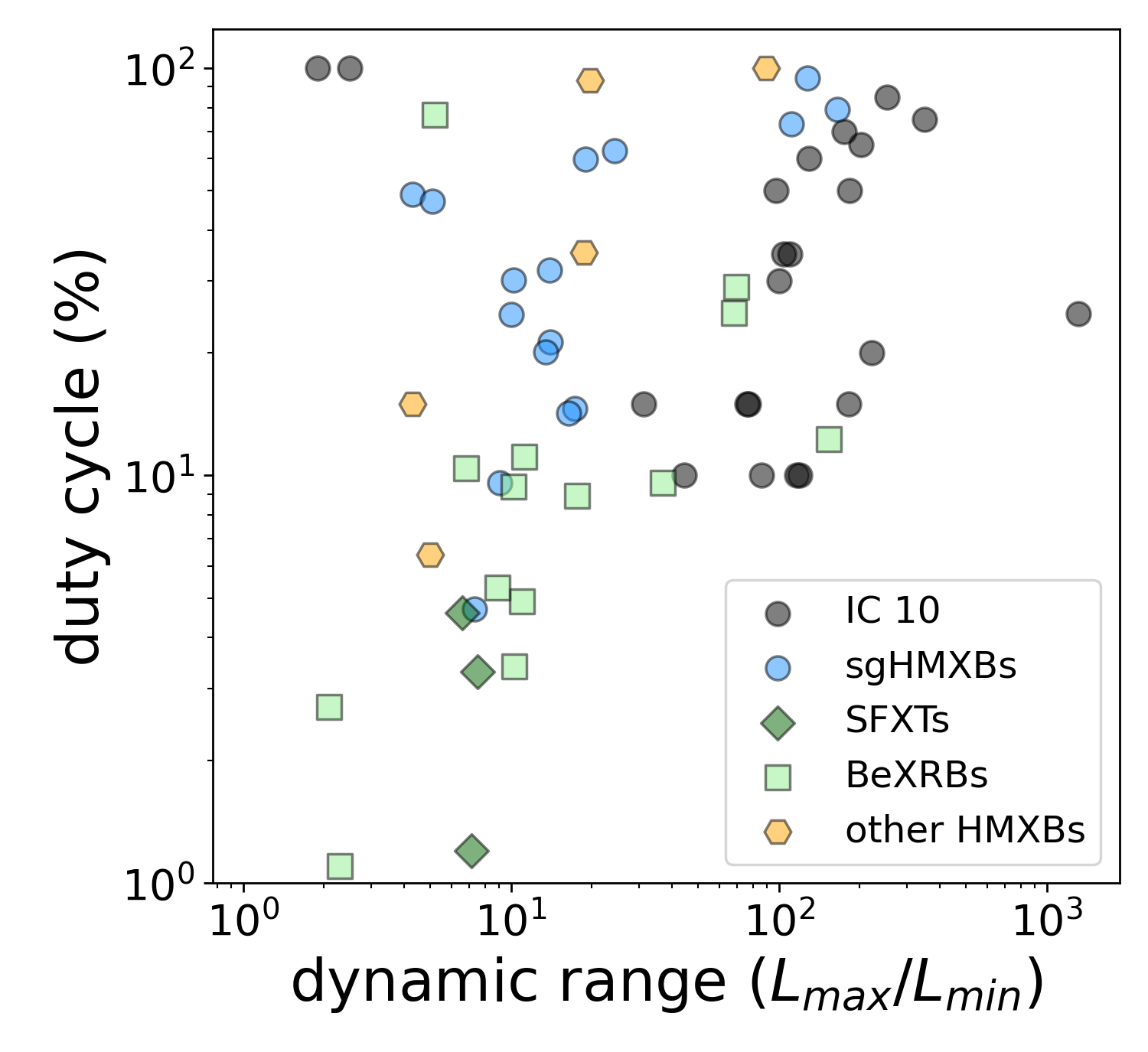}
    \caption{The DC as a function of DR for IC~10 HMXB candidates (black circles), compared to various subclasses of Galactic HMXBs \citep[][blue, green, and yellow symbols]{Sidoli+18}.}
    \label{fig:sidoli_compare}
\end{figure}

Table~\ref{tab:hmxbs} summarizes the long-term and rapid variability properties of the IC 10 HMXB candidates, including the number of observations in which this null hypothesis of a constant CR can be rejected or not rejected, the observed DR, constraints on the DC, and the peak \Lx\ observed for each source. In Figure~\ref{fig:sidoli_compare}, we compare the observed best estimate DCs and DRs derived for each IC~10 HMXB candidate to the Galactic XRB sample described by \citet{Sidoli+18}. \citet{Laycock+17bsg} found an unusually high fraction of HMXB candidates with blue supergiant counterparts, and we find that the IC~10 HMXB candidates exhibit similar DCs as Galactic sgHMXBs, although the DR of the IC~10 sample is roughly an order of magnitude larger than for the Galactic sample. Although our analysis is not sensitive to very low DCs ($<$1\%), we do not observe a significant pileup of sources with DCs$<$5-10\%, as might be expected if IC~10 hosted a large population of BeXRBs. Both theoretical and observational studies suggest that younger stellar populations and lower metallicities are associated with brighter-\Lx\ HMXBs \citep{Linden+10,Fragos+13a,Fragos+13b,Lehmer+19,Lehmer+22}; thus, the larger peak outburst luminosities relative to the baseline ``quiescent'' luminosity of HMXBs in IC~10 may potentially be due to the systematically younger ages of the HMXBs in IC~10, the lower metallicity of the galaxy ($\sim$0.2\Zsun), or both. The relatively low peak luminosities observed in the IC~10 HMXBs ($\sim$10$^{36-37}$ \lum) suggests that Type I bursts are responsible for most of the X-ray variability observed in the IC~10 population.

\section{The X-ray Luminosity Function and Star Formation History of IC~10}\label{sec:xlf}
There is a considerable body of work demonstrating the link between the X-ray luminosity functions (XLFs) of XRB populations and properties of the host galaxy, such as stellar mass, star formation rate, and metallicity \citep{Grimm+03,Gilfanov04,Gilfanov+04,Kim+04,Zhang+12,Mineo+12,Lehmer+19,Lehmer+22,Lehmer+24}. With a relatively clean (i.e., largely free from foreground contaminating sources) sample of HMXB candidates intrinsic to IC~10, we can construct the XLF of IC~10's HMXB population for the first time. Since IC~10 is commonly classified as a starburst galaxy, we expect the HMXB population to be dominated by young and massive HMXBs. Tracers of very recent star formation (the last few tens of Myr) yield high SFRs; e.g., the H$\alpha$ flux implies a SFR of $\sim$0.2-0.6 \Msun\ yr$^{-1}$ depending on the extinction value assumed \citep{Leroy+06}, but observations of the Wolf-Rayet (WR) population in IC~10 \citep{Crowther+03} and the radio continuum suggest a SFR of $\sim$0.02-0.05 \Msun\ yr$^{-1}$ \citep{Tehrani+17,Basu+17}. The older and intermediate-aged populations (with ages larger than a few Gyr), such as those traced by planetary nebulae, red horizontal branch stars, and carbon stars, suggest significantly lower SFRs, on the order of $\sim$0.02 \Msun\ yr$^{-1}$ \citep{Magrini+03,Demers+04}.

\citet{Weisz+14} constructed the SFH of IC~10 and other nearby dwarf galaxies using {\em Hubble} WFPC2 photometry of the resolved stellar populations. However, this study specifically focused on the SFHs during the epoch of reionization; further work by \citet{DellAgli+18}, which used infrared observations of asymptotic giant branch and red supergiant stars, was able to refine the SFH of IC~10 in the last few Gyr. Additionally, \citet{DellAgli+18} explicitly incorporated the metallicity evolution of IC~10 into their models \citep{Yin+10}. This SFH, however, underpredicts the high SFRs in the last $\sim$6-10 Myr predicted by observations of the H$\alpha$ flux, radio continuum, and WR population in IC~10 by more than an order of magnitude.

\begin{figure*}
    \centering
    \begin{tabular}{cc}
        \includegraphics[width=0.5\linewidth]{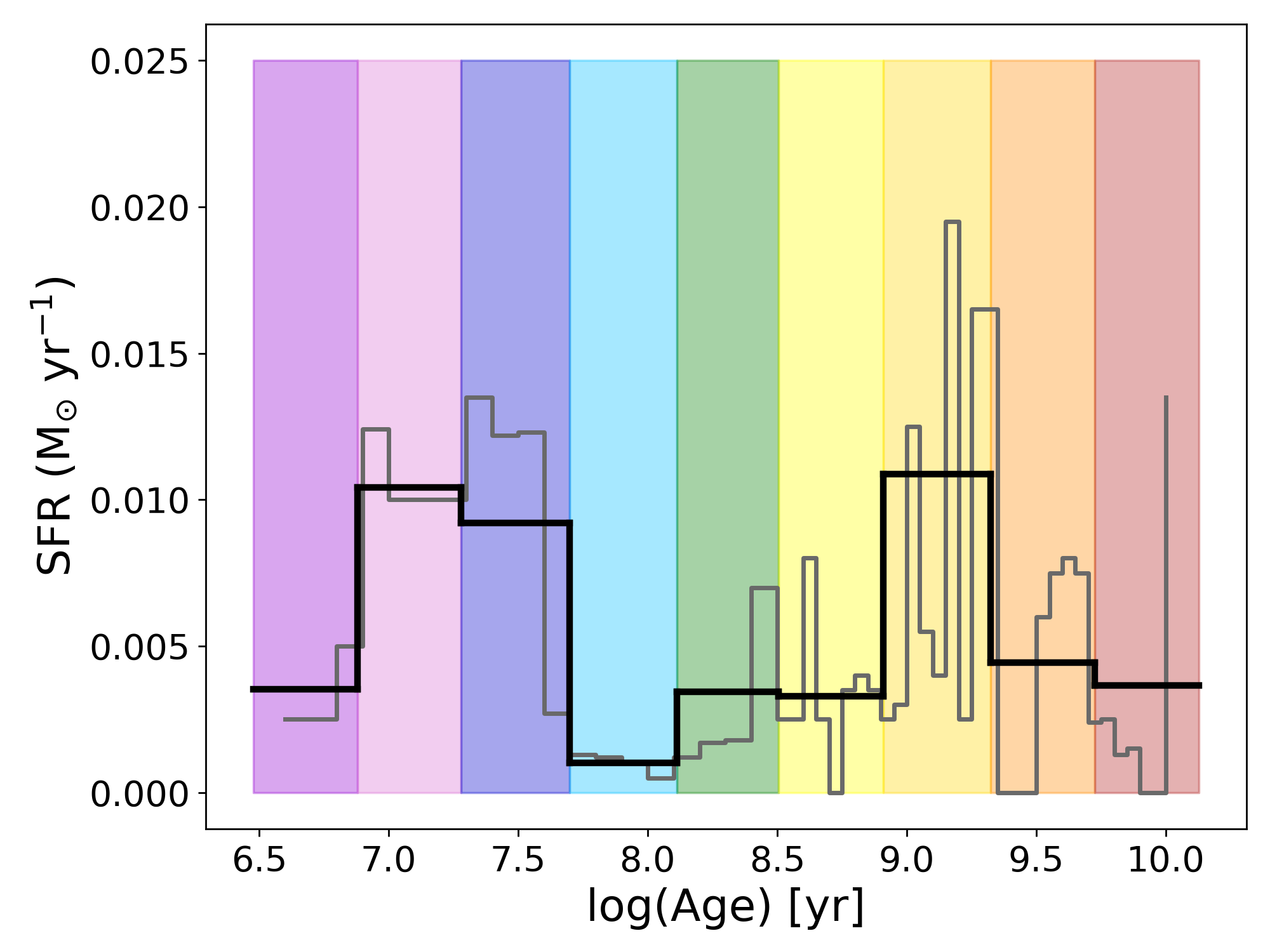} &
        \includegraphics[width=0.5\linewidth]{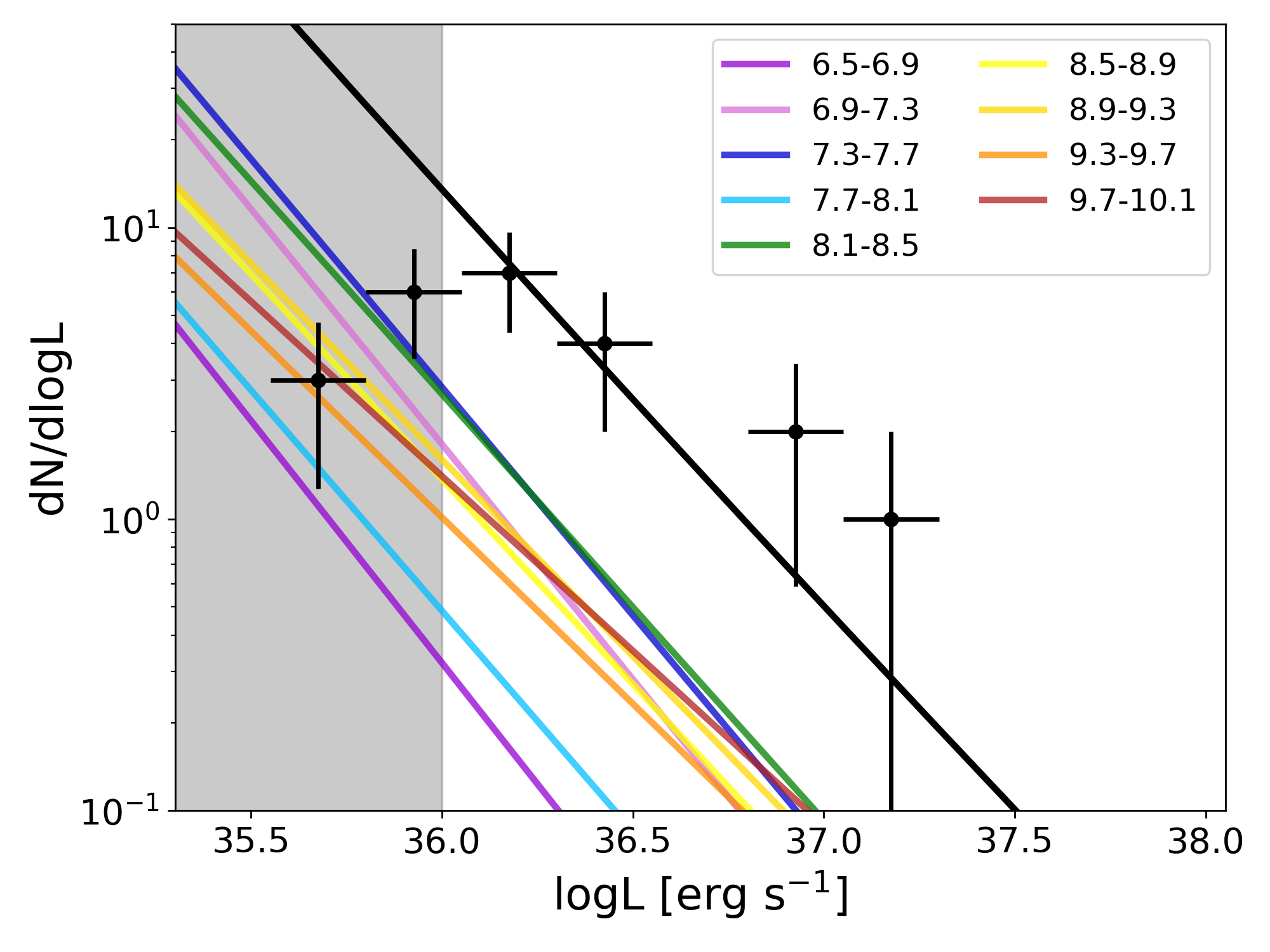} \\
    \end{tabular}
    \caption{{\it Left}: The SFH from \citet[][gray line]{DellAgli+18}, with the temporal bins used to construct XLF basis functions from \citet{Lehmer+24} indicated with various colors. The average SFR in each temporal bin (i.e., the SFH adopted in this analysis) is shown by the thick black line. {\it Right}: The observed XLF of IC~10 (black points) and the XLF components predicted by the \citet{Lehmer+24} XLF basis functions (colors correspond to the temporal bins shown in the left panel; the logarithms of the ages is given in the legend). The thick black line indicates the total predicted XLF of the IC~10 XRB population. The gray shaded region indicates the typical detection limit of a single observation.}
    \label{fig:litSFH}
\end{figure*}

The XLF of IC~10 can provide an additional constraint on the very recent SFR of the galaxy. Recently, \citet{Lehmer+24} constructed XLF ``basis functions'' which can be used to predict the XLF of a galaxy's XRB population based on the galaxy's SFH and metallicity. In their empirical framework, the XLF of a galaxy's XRB population is described as

\begin{equation}
    \frac{dN}{d\text{log}L} = \sum^{n_{\rm SFH}}_{j=1} M_{\star}(t_j) \frac{dN(t_j,Z_j)}{d\text{log}L dM_{\star}},
\end{equation}

\noindent where $n_{\rm SFH}$ is the number of bins used to construct the SFH (in this case, $n_{\rm SFH}=9$), $M_{\star}(t_j)$ is the stellar mass formed in the $j$th time bin, and $Z$ is the metallicity in the $j$th time bin. The SFH and metallicity evolution from \citet{DellAgli+18}, which is reported in units of \Msun\ yr$^{-1}$ (given in Table~\ref{tab:sfh}), is used to calculate $M_{\star}(t)$, and expressions for the metallicity- and time-dependent functions $dN/d$log$L$ are given in \citet{Lehmer+24}. We adopt the time bins used by \citet{Lehmer+24}, which are equally spaced in log time by 0.4 dex, and calculate the time-weighted average SFR from \citet{DellAgli+18} in each bin. The input SFH, resulting XLF predictions, and the observed XLF of IC~10 are shown in Figure~\ref{fig:litSFH}. To construct the observed (differential) XLF, we used the brightest peak \Lx\ of each source and luminosity bins of $\Delta$log\Lx\ of 0.25. In general, we find that the peak \Lx\ of each source exceeds the mean \Lx\ by $\sim$45\%, which corresponds to a shift in log\Lx\ of $\sim$0.16. The assumption of peak \Lx, rather than mean \Lx, is therefore unlikely to have a significant impact on the overall shape of the XLF. We include IC~10 X-1 in the XLF assuming a typical peak \Lx\ of $\sim$7$\times10^{37}$ \lum\ \citep{Laycock+15a,Laycock+15b}. We used the Gehrels approximation \citep{Gehrels86} to estimate the upper and lower bounds on the differential number of HMXBs per luminosity bin. The turnover in the XLF at log\Lx$\sim$36 is due to incompleteness in the individual \Chandra\ exposures. While the predicted XLF agrees with the observed number of fainter sources ($\sim$1-3$\times10^{36}$ \lum), there is an excess of sources brighter than $\sim$6$\times10^{36}$ \lum.

\begin{figure}
    \includegraphics[width=1\linewidth]{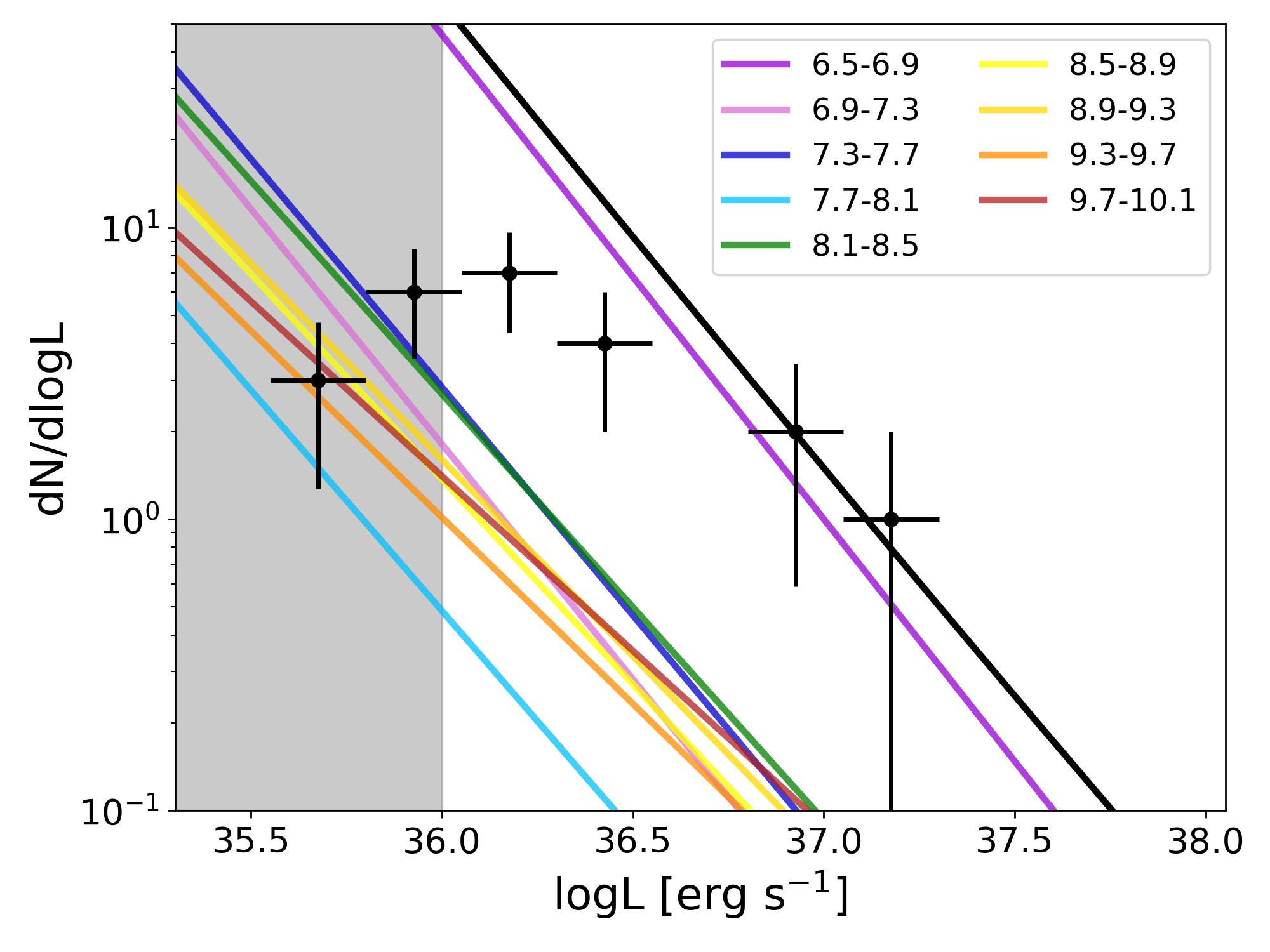}
    \caption{Same as the right panel of Figure~\ref{fig:litSFH}, but assuming a recent ($\lesssim$7.6 Myr) SFR of $\sim$0.5 \Msun\ yr$^{-1}$.}
    \label{fig:XLF_recentSFR}
\end{figure}

We experimented with adjusting the SFR in the 3.0-7.6 Myr bin to match various values reported in the literature and found that a recent SFR of $\sim$0.5 \Msun\ yr$^{-1}$ (with a metallicity of 0.2\Zsun) adequately describes the bright end of the observed XLF, as shown in Figure~\ref{fig:XLF_recentSFR}, although it significantly over-predicts the number of fainter sources. If such a burst of star formation did occur $\lesssim$8 Myr, it is likely we are only just beginning to observe the most massive binary systems evolve into HMXBs; these ages are young enough that the more massive O-type donors to HMXBs could still be in the main sequence phase or have just evolved into a post-main sequence, supergiant phase of their evolution. The lower mass systems (which would produce systematically less luminous HMXBs) would have not yet undergone compact object formation and entered the HMXB phase. Furthermore, the overall low luminosities and long-term variability of HMXBs in IC~10 (no sources observed with \Lx$\gtrsim10^{38}$ \lum) suggests that the population is dominated by wind-fed systems, rather than the typically brighter and less variable Roche lobe overflow-dominated lower mass systems. 

\begin{table}
\centering
\caption{Adopted Star Formation History of IC 10}\label{tab:sfh}
    \begin{tabular}{ccc}
    \hline \hline
    Age Bin   & SFR                & $Z$ \\
    (Myr)   & (\Msun\ yr$^{-1}$) & ($Z_{\odot}$)  \\
    (1) & (2) & (3) \\
    \hline
    3.0-7.6      & 0.0035 & 0.20 \\  
    7.6-19       & 0.0104 & 0.20 \\
    19-50        & 0.0092 & 0.20 \\
    50-130       & 0.0010 & 0.20 \\
    150-320      & 0.0034 & 0.10 \\
    320-810      & 0.0033 & 0.10 \\
    810-2100     & 0.0101 & 0.05 \\
    2100-5300    & 0.0044 & 0.05 \\
    5300-13400   & 0.0037 & 0.05 \\
    \hline \hline
    \end{tabular}
\end{table}

\section{Discussion and Conclusions}\label{sec:discussion}
Our analysis of the X-ray variability and the XLF of strong HMXB candidates in IC~10 reveals a population of X-ray sources dominated by young sgHMXB systems, consistent with inferences from previous work \citep{Laycock+17bsg,Laycock+17var}. The prevalence of supergiant donors stands in contrast to the HMXB populations of comparable nearby galaxies, such as the SMC, which are dominated by BeXRBs \citep{Antoniou+10,Ducci+14}. This difference in the dominant donor star type is likely related to the time period of maximum star formation of the host galaxy. While the peak star formation in the SMC occurred $\sim$40 Myr ago \citep[the typical lifetime of a B-type main sequence star;][]{Antoniou+10,Antoniou+19}, there is evidence -- including the XLF of HMXBs in IC~10 presented in this work -- that the SFR of IC~10 has risen dramatically in the past $\sim$8 Myr. The overall young age of the HMXB population in IC~10 also suggests that the observed HMXBs reside much nearer to their original birthplaces than in other galaxies \citep{Binder+23}. We note that the HMXB population of IC~10 is overall significantly fainter than the HMXB populations used to construct XLF basis functions of \citet{Lehmer+24}, and we have not formally accounted for uncertainties in these basis functions in our analysis. While we cannot directly constrain the very recent SFR of IC~10 through the XLF alone, we can qualitatively say that the observed XLF of IC~10 HMXBs is consistent with IC~10 having recently experienced a star formation event, the timing and magnitude of which is within the ranges measured via independent multiwavelength studies. Further extension of the \citet{Lehmer+24} XLF basis functions to lower \Lx\ are needed to more rigorously compare to the IC~10 HMXB population.

Accretion by a compact object from a clumpy stellar wind is thought to be the primary mechanism by which X-rays from sgHMXBs are produced. This model predicts a modest baseline in \Lx\ ($\sim$10$^{36-37}$ \lum) and significant X-ray variability \citep{Oskinova+12,MartinezNunez+17}, which is observed in the IC~10 HMXB candidates. We find that lower-flux HMXB candidates exhibit lower DCs than brighter sources. The XLF basis functions of \citet{Lehmer+24} were derived from empirical modeling of the XLFs of bright XRBs across a wide range of galaxies, most of which exhibit high SFRs and/or high stellar masses, and samples a range in \Lx\ from $\sim$10$^{37}$ \lum\ to $\sim$10$^{40}$ \lum. That these XLF basis functions can describe the observed XLF in IC~10, which is systematically of lower luminosity, and provide insight into the recent SFR of IC~10 is remarkable. However, we stress that the XLF of IC~10 presented in Section~\ref{sec:xlf} uses the maximum observed \Lx\ across a $\sim$19 year baseline for each of the detected sources. Utilizing only a single observation -- even if that observation had an exposure time of $\sim$235 ks, equivalent to the combined exposure time of all IC~10 observations considered here -- would (1) fail to capture the entire HMXB population, as many systems would be in an ``off'' state during the observation, which would decrease the normalization of the XLF, and (2) capture many ``on'' systems at less than their peak-\Lx, which would alter the slope of the XLF. Understanding the correlation between X-ray variability and luminosity in faint ($\lesssim10^{36}$ \lum) HMXBs is thus important for correct interpretation of the XLF and its relationship to SFH and metallicity, particularly as advances in X-ray detector sensitivity and effective area promise to capture observations of faint sources efficiently in modest exposure times \citep[such as with the {\em Advanced X-ray Imaging Satellite, AXIS;}][]{SafiHarb+23}.

\begin{acknowledgments}
B. A. B. acknowledges support from HST-GO-16769 and the National Science Foundation Launching Early-Career Academic Pathways in the Mathematical and Physical Sciences (LEAPS-MPS) award \#2213230. This research has made use of data obtained from the \Chandra\ Data Archive and software provided by the \Chandra\ X-ray Center (CXC) in the application package CIAO. S. L. and S. B. acknowledge support of  NSF Astronomy and Astrophysics  grant \# 2109004. The authors thank Rebecca Kyer and Shelby Albrecht for their assistance with constraining source DCs. This paper employs a list of \Chandra\ datasets, obtained by the \Chandra\ X-ray Observatory, contained in~\dataset[doi:10.25574/cdc.428]{https://doi.org/10.25574/cdc.428}.
\end{acknowledgments}

\vspace{5mm}
\facilities{CXO}

\software{astropy \citep{astropy:2013,astropy:2018,astropy22}, CIAO \citep{Fruscione+06}}

\bibliography{sample631}{}
\bibliographystyle{aasjournal}

\end{document}